\documentclass[iop]{emulateapj}
\usepackage{amsmath}

\begin{document}

\title{Dark Matter Admixed Type Ia supernovae}

\author{S.-C. Leung\thanks{Email address: scleung@phy.cuhk.edu.hk}, 
M.-C. Chu\thanks{Email address: mcchu@phy.cuhk.edu.hk}, and 
L.-M. Lin\thanks{Email address: lmlin@phy.cuhk.edu.hk}
}

\affiliation{Department of Physics and Institute 
of Theoretical Physics, The Chinese University 
of Hong Kong, Hong Kong S.A.R., China}

\date{\today}

\begin{abstract}

We perform two-dimensional hydrodynamic simulations for the thermonuclear 
explosion of Chandrasekhar-mass white dwarfs with dark matter (DM) cores in 
Newtonian gravity. We include a 19-isotope nuclear reaction network and make 
use of the pure turbulent deflagration model as the explosion mechanism in 
our simulations. Our numerical results show that the general properties
of the explosion depend quite sensitively on the mass of the DM core $M_{\rm DM}$:
a larger $M_{{\rm DM}}$ generally leads to a weaker 
explosion and a lower mass of synthesized iron-peaked elements. 
In particular, the total mass of ${}^{56}$Ni produced can drop from about
$0.3$ to $0.03 M_\odot$ as $M_{\rm DM}$ increases from 0.01 to 
$0.03 M_\odot$. 
We have also constructed the bolometric light curves 
obtained from our simulations and found that our results match well with
the observational data of sub-luminous Type-Ia supernovae.

\end{abstract}


\maketitle

\section{Introduction}
\label{sec:intro}

\subsection{Type-Ia supernovae}

Type-Ia supernovae (SNIa) are important astrophysical objects because of 
the similarity in their light curves and spectra
\citep{Branch1992}, which leads to wide applications 
of SNIa in cosmological distance measurement such as the determination of 
the Hubble parameters \citep{Leibundgut1992} and the discovery of the 
accelerating expansion of the universe \citep{Riess1998, Perlmutter1999}.  
However, despite their important roles in modern cosmology, both the 
progenitor system and explosion mechanism of SNIa are not yet fully 
understood. 
While it is generally believed that SNIa are due to the thermonuclear 
explosion of a carbon-oxygen white dwarf (WD) in binary systems, it is still 
unclear whether the companion is a normal non-degenerate star or another WD. 
Traditionally, SNIa is attributed to the explosion 
of a WD at the Chandrasekhar mass limit \citep{Arnett1969}.
The WD has a mass initially far from the mass limit. 
Depending on the accretion rate, the mass can either gradually grow 
until the baryonic matter becomes degenerate \citep{Nomoto1982a},
or a detonation front forms at the envelope and sheds away the outer 
mass \citep{Nomoto1982b}. Both mechanisms provide conditions for the 
formation of a first trigger \citep{Nomoto1984} which spreads in the form of 
a deflagration wave \citep{Nomoto1976} and unbinds the star. 
However, neither pure deflagration \citep{Nomoto1984} nor pure detonation 
model \citep{Arnett1969} is adequate to explain the observed velocity 
profile, optical light curve, spectra, galactic chemical abundance and 
explosion strength. Furthermore, recent studies show that SNIa 
can be formed without invoking Chandrasekhar mass WD \citep{Scalzo2014}.
For example, violent white dwarf mergers  
can also explain the SNIa distributions \citep{Pakmor2013}. 

The difficulties encountered by
the pure deflagration and pure detonation models have led
to extensions of models including the pure turbulent 
deflagration (PTD) model 
\citep{Reinecke1999b, Reinecke1999a, Reinecke2002a, Reinecke2002b},
delayed-detonation transition (DDT) model 
\citep{Khokhlov1989, Khokhlov1991a, Khokhlov1991b,
Khokhlov1991c, Khokhlov1997} and gravitationally confined detonation (GCD)
model (previously known as the detonation failed deflagration 
model)
\citep{Plewa2007, Kasen2007, Jordan2008, Meakin2009, Jordan2012}. 
Each model has its own theoretical difficulties. For example, 
while the PTD model can produce explosion with a variety of 
strengths \citep{Roepke2006}, there are still unburnt 
low-velocity carbon and oxygen near the core, which are not observed 
\citep{Roepke2007b}. The DDT model can provide sufficient 
intermediate mass elements (IME) and leave very little fuel 
\citep{Gamezo2004, Gamezo2005}. However, the possibility of transition is 
still being debated \citep{Lisewski2000}.

In recent years, the PTD, DDT and GCD models are studied extensively in 
multi-dimensional simulations \citep{Long2014, Seitenzahl2013, Jordan2012}. 
The models can well explain the phenomena of normal SNIa, i.e., supernovae
with a correlated peak luminosity against B-band decline rate, chemical 
stratification and a large velocity gradient \citep{Benetti2005}.
However, there is a significant number of peculiar SNIa which 
are sub-luminous and super-luminous \citep{Li2001}. 
In particular, sub-luminous SNIa have a much lower absolute magnitude of 
B-band at maximum. For example, the famous SN1991bg \citep{Filippenko1992} 
recorded a 2.5 mag and 1.6 mag dimmer in B- and V-band peak magnitudes. 
The B-band decline rate is faster than the norm, with a 
lower expansion velocity and stronger Si II absorption lines
\citep{Doull2011}. It was initially assumed that such unusual SNIa are 
extremely rare. 
However, there are now sufficient number of sub-luminous SNIa that they are  
classified as the FAINT group as suggested in \citep{Benetti2005}. Detailed 
study shows that the light curves in this group of SNIa are 
also homogeneous among themselves as those of normal SNIa
\citep{Doull2011}. 
The faintest SNIa ever found to date is SN2008ha
\citep{Foley2009}, with a low magnitude of $M_{V} = -14.2$ mag and 
extremely low expansion velocity $\sim 2000$ km s$^{-1}$.

In view of the discovery of sub-luminous and super-luminous SNIa, 
the explosion of a Chandrasekhar-mass WD can no longer be the sole 
explanation of SNIa because of the lack of variety in its explosion. 
The sub-Chandrasekhar mass double detonation model is often 
regarded as the explanation for sub-luminous SNIa \citep{Woosley1994}. 
The model suggests that when the mass accretion of a WD from its
companion is adequately fast, the matter on its envelope,
mostly helium, can be ignited and an implosion is triggered 
\citep{Nomoto1982b}. The front converges at the WD core and a second
explosion is created. 
This model allows a WD to be burnt if the matter is not yet degenerate. 
By tuning the host WD mass, less luminous SNIa can be modeled, which can be 
fitted to explain certain sub-luminous SNIa, 
for instance SN1991bg \citep{RuizLapuente1993}.
The helium detonation is found robust in inducing a second explosion
\citep{Fink2009} and the predicted optical signal is compatible with 
observations \citep{Kromer2010, Sim2012}. However, recent studies of 
this model with less massive helium shell show similar features
as normal SNIa \citep{Sim2010, Ruiter2011, Ruiter2014} instead
of sub-luminous ones. Furthermore, the detonation might not be
started robustly \citep{Livne1990b}. 
Even when a helium detonation is triggered, the detonation wave might not 
penetrate deep into the carbon/oxygen core \citep{Moll2013}, and the 
distribution of outer chemical elements can be in conflict with observation 
data \citep{Hoeflich1996a, Hoeflich1996b}. 

Another popular proposal for explaining 
sub-luminous SNIa is the pure turbulent deflagration model with remnant. 
This model assumes that the deflagration only partially
burns the WD, and parts of the WD remain bounded after
the explosion. It is applied to the SN2002cx
class of the sub-luminous SNIa. The 
synthetic color light curves and the spectra
can match well with the observational 
data \citep{Jordan2012,Kromer2013,Fink2014}.
In \citep{Kromer2015} this model is further shown 
to be in good agreement with the faintest SN2008ha.
The recent observational hints of the SN2008ha 
remnant \citep{Foley2014} also support this model
as the origin of this class of sub-luminous SNIa.

Violent merging of two low-mass WD's is also a 
possible candidate of explaining the sub-luminous
SNIa. For example, \citep{Pakmor2010} showed
that this model provides a good match to the 
observed features of SN1991bg, while in 
\citep{Kromer2013b} the light curves and spectra
of SN2010Ip are well reproduced.


\subsection{Dark matter astrophysics}

The effects of various dark matter (DM) 
candidates on stellar evolution and structure have 
been studied in details. 
For example, the effects of DM annihilation as the energy source in 
early stars have been considered in \citep{Spolyar2008, Ripamonti2010,
Fairbairn2008,Freese2009,Spolyar2009,Hirano2011}. 
The DM particle capture and evaporation rates of the sun 
\citep{Gould1987a, Gould1987b} and the Earth \citep{Gould1988} were studied
in early 1990s. 
The dense core of compact stars is also a good probe of DM 
\citep{Bertone2008, Fan2011, Lavallaz2010}.
DM particles can annihilate or decay 
inside a compact star and thus provide an energy source
\citep{Gonzalez2010, PerezGarcia2014}. For example, it has been suggested 
that the energy is sufficient to maintain the surface temperature of WDs 
\citep{Moskalenko2007, Hooper2012} and neutron stars
\citep{Kouvaris2008, Kouvaris2010}. 
On the other hand, non-self-annihilating DM can affect the star by its gravity. 
The self-gravitating DM core inside a compact star might 
collapse, which forms a black hole and 
engulfs the star \citep{Goldman1989, Lavallaz2010}. 
The detection of ancient compact stars can thus provide limits on the DM 
scattering cross section for different types of DM particles
\citep{Kouvaris2011a,Kouvaris2012,McDermott2012,Bramanta2013}.



\subsection{Motivation}

Previously, we have studied the 
equilibrium structure and stability of compact 
stars with cores composed of non-self-annihilating DM 
particles which are modeled by an ideal Fermi gas 
\citep{Leung2011,Leung2012,Leung2013}. In 
particular, for DM particle mass of about 1 
GeV, we found that the DM core 
can affect the structure of a WD significantly. 
The DM core can be as massive as about $0.01M_\odot$ 
and the Chandrasekhar mass limits of these WD 
can be smaller than those without DM by as much as 
40\%. An implication of our findings in \citep{Leung2013} 
is that the initial conditions of SNIa might not 
be as universal as generally assumed. 
In this paper, we study how DM affects SNIa explosions by 
performing two-dimensional hydrodynamic simulations of the 
thermonuclear explosions of Chandrasekhar-mass WDs with DM 
cores. We find that these objects generally have weaker 
explosions and lower masses of synthesized iron-peaked 
elements, and hence they may account 
for the sub-luminous class of SNIa.  

The plan of this paper is as follows: In Section \ref{sec:methods} we outline
the equations and methods that we used in the numerical simulations. 
Section \ref{sec:results} presents the general results of our SNIa
simulations in terms of the explosion energy, 
nucleosynthesis, and features of the propagating flame surface. 
We also compare the bolometric light curves constructed from our simulations
with the observational data of sub-luminous SNIa. 
Finally, we summarize in section \ref{sec:conclude}.

\section{Methods}
\label{sec:methods}

We have developed a two-dimensional hydrodynamical code with Newtonian gravity 
to model SNIa. 
The code makes use of the Weighted Essential Non-Oscillatory (WENO) scheme
for spatial discretization \citep{Barth1999}. This is a fifth-order scheme 
which processes piecewise smooth functions with discontinuities in order to 
simulate the flux across grid cells with high precision, while avoiding 
spurious oscillations around the shock. 
The discretization in time is performed by using the five-stage, third-order, 
non-strong stability preserving explicit Runge-Kutta scheme \citep{Barth1999}. 
Various consistency and convergence tests 
have been done to validate our code \citep{Leung2015a}. 
Here we only outline the essence of the code. 

\subsection{Initial Model}

In the simulation there are both baryonic normal matter (NM)
and DM. The initial density profiles are obtained by 
solving the hydrostatic equilibrium equations for both NM and DM: 
\begin{eqnarray}
\frac{dp_{{\rm NM}}}{dr} = -\frac{G (M_{c({\rm NM})}(r) + M_{c({\rm DM})}(r))}{r^2} \rho_{{\rm NM}}, \\
\frac{dp_{{\rm DM}}}{dr} = -\frac{G (M_{c({\rm NM})}(r) + M_{c({\rm DM})}(r))}{r^2} \rho_{{\rm DM}} ,
\end{eqnarray}
where $\rho_{{\rm NM}}$ and $\rho_{{\rm DM}}$ are the NM and DM density, 
respectively. The enclosed masses of $M_{c({\rm NM})}$ and $M_{c({\rm DM})}$ are
determined by 
\begin{eqnarray}
\frac{dM_{c({\rm NM})}}{dr} = 4 \pi r^2 \rho_{\rm NM} , \\
\frac{dM_{c({\rm DM})}}{dr} = 4 \pi r^2 \rho_{\rm DM} .
\end{eqnarray}
The initial NM is assumed to be isothermal with a temperature of $10^8$ K. 
The chemical composition is 50\% $^{12}$C and 50\% $^{16}$O by mass. 
To construct the initial WD models and simulate the NM dynamics,
we employ the equation of state (EOS) developed
and calibrated in \citep{Timmes1999b, Timmes1999c}.
The EOS describes the equilibrium thermodynamics properties of a gas
which includes 1. electrons in the form of an ideal gas with arbitrarily 
degenerate and relativistic levels, 2. ions in the form of a classical ideal 
gas, 3. photons described by the Planck distribution,
4. contributions from electron-positron pairs. On the other hand, 
the DM is modeled by an ideal degenerate Fermi gas with a particle mass $m_{{\rm DM}}$ of
1 - 10 GeV \citep{Leung2011,Leung2012,Leung2013}, which is motivated by recent 
hints on possible detection of GeV scale DM particles in 
some direct DM searches, such as DAMA and CoGeNT 
\citep{Bernabei2013, Aalseth2011}. 
One concern is that the admixed DM core 
may alter the stellar evolution path 
during the main-sequence stage, where deviations 
from standard stellar evolution theory have 
not been observed. We show in the Appendix that the range 
of $M_{{\rm DM}}$ considered does not bring significant 
effects on the evolution tracks, especially of
the carbon-oxygen WD progenitors, namely 
4 - 7 $M_{\odot}$ main-sequence stars.

\begin{figure}
\centering
\includegraphics*[width=8cm, height=6cm]{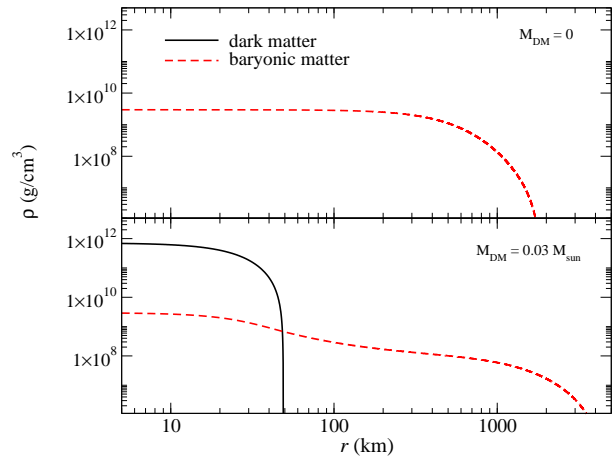}
\caption{Upper panel: initial NM density profile of Model 2D-PTD-3-0-c3
listed in Table~\ref{table:TestPTD}. 
Lower panel: initial NM (dashed line) and DM (solid line) density profiles 
of Model 2D-PTD-3-3-c3.}
\label{fig:rho_plot}
\end{figure}

We plot in Fig. \ref{fig:rho_plot} the density profiles of two of our 
initial models with $m_{{\rm DM}} = 1$ GeV. The 
upper panel of Fig.~\ref{fig:rho_plot} is a standard 
model without DM (Model 2D-PTD-3-0-c3 in Table~\ref{table:TestPTD}).
The lower panel shows the NM and DM profiles of a stellar model with a DM 
core of mass $M_{\rm DM}=0.03 M_\odot$ (Model 2D-PTD-3-3-c3 in 
Table~\ref{table:TestPTD}). 
The central DM density of this model is about 3 orders of magnitude higher 
than that of the NM. Due to its high compactness, the DM core 
produces a strong gravitational field around the core region and leads to a 
cusp-like structure in NM and steep NM density gradient in the core. 
We remark that the density profiles shown here are 
different from those in \citep{Leung2013} owing to the choice
of DM particle mass and the total mass of admixed 
DM. In Fig. 6 of \citep{Leung2013} the EOS of DM particle
is chosen to be 10 GeV ideal degenerate Fermi gas instead of
1 GeV in this article. Also, in the same figure
of \citep{Leung2013}, the admixed DM mass is about 
$10^{-3} M_{\odot}$, which is one order of magnitude 
lower than those in Fig. \ref{fig:rho_plot}.

\subsection{Hydrodynamics}

The thermonuclear explosion of a WD with a DM core is inherently 
a two-fluid system where the NM and DM couple through gravity. 
In principle, one has to model the dynamics of the two fluids consistently 
by solving two different sets of hydrodynamics equations.
However, the typical density of DM in our simulations is about two or three 
orders of magnitude higher than that of NM. The size of the DM core is also 
much smaller than the stellar radius. As a result, the dynamical time and
length scales of NM and DM differ by orders of magnitude, and hence 
performing a consistent and accurate two-fluid hydrodynamics simulation for 
SNIa would be a computationally challenging task.  
On the other hand, due to its high compactness, the dynamics of 
the DM core is governed mainly by its self-gravity. 
The motion of NM near the core is influenced 
by the DM, but not vice versa. Furthermore, the total energy release and 
nucleosynthesis in the explosion depend mainly on the propagation 
of the flame, which lies well outside the DM core. 
It may thus be reasonable to neglect the motion of DM in the explosion.

As a first step towards understanding the effects of DM on SNIa explosions, 
we only model the dynamics of NM in the simulations. The DM core
is assumed to be stationary and affects the NM only through its 
gravitational field. 
The hydrodynamic code solves the two-dimensional Euler equations for NM 
in cylindrical coordinates $(r, z)$ with a detailed nuclear reaction network 
coupled with sub-grid turbulence. The equations are
\begin{flalign}
\frac{\partial \rho_{{\rm NM}}}{\partial t} + \nabla \cdot (\rho \vec{v}) = 0,
\label{eq:euler_mass} \\
\frac{\partial \left( \rho_{{\rm NM}} \vec{v} \right)}{\partial t} + 
\nabla \cdot (\rho \vec{v} \vec{v}) = -\nabla P - \rho \nabla \Phi,
\label{eq:euler_momentum} \\
\frac{\partial \tau}{\partial t} + \nabla \cdot [\vec{v} (\tau + p)] = \rho \vec{v} \cdot \nabla \Phi
+ Q_{{\rm nuc}} - Q_{{\rm turb}} - Q_{\nu},  
\label{eq:euler_energy} \\
\frac{\partial \left( \rho_{{\rm NM}} q \right)} {\partial t} 
+ \nabla \cdot \left( \rho_{{\rm NM}} q \vec{v} \right) =
Q_{{\rm turb}} + \nabla \cdot (\rho_{{\rm NM}} \nu_{\rm turb} \nabla q) , 
\label{eq:euler_sgs}
\end{flalign}
where $\rho_{{\rm NM}}$, $v_r$, $v_z$, $p_{{\rm NM}}$ and $\tau$ are the mass
density, velocities in the $r$ and $z$ directions, pressure 
and total energy density of the baryonic matter. The total energy
density includes both the thermal and kinetic contributions
$\tau = \rho_{{\rm NM}} \epsilon + \frac{1}{2} \rho_{{\rm NM}} v^2$, 
where $\epsilon$ is the specific internal energy. The specific turbulence 
energy $q$ is determined by Eq.~(\ref{eq:euler_sgs}).
The gravitational potential $\Phi$ is sourced by both fluids and is 
determined by the Poisson equation
\begin{equation}
\nabla^2 \Phi = 4 \pi G (\rho_{{\rm NM}} + \rho_{{\rm DM}}) . 
\end{equation}

In Eqs.~(\ref{eq:euler_energy}) and (\ref{eq:euler_sgs}), $Q_{{\rm nuc}}$ 
and $Q_{\rm turb}$ are the heat sources from nuclear fusions and sub-grid 
turbulence, respectively; $Q_{\nu}$ is the heat loss due to neutrino emission, and 
$\nu_{\rm turb}$ is the effective eddy viscosity. We refer the reader to 
\citep{Niemeyer1995a, Reinecke2002a} for a detailed discussion on how these
quantities are determined in the simulations. 
To calculate the heat production from nuclear reactions, we incorporate
the 19-isotope nuclear reaction network subroutine developed by 
\citet{Timmes1999a} into our hydrodynamic code. 
The isotopes include $^{1}$H, $^{3}$He, $^{4}$He, 
$^{12}$C, $^{14}$N, $^{16}$O, $^{20}$Ne, $^{24}$Mg, 
$^{28}$Si, $^{32}$S, $^{36}$Ar, $^{40}$Ca, $^{44}$Ti,
$^{48}$Cr, $^{52}$Fe, $^{54}$Fe, $^{56}$Ni, neutron
and proton. The fusion network includes reactions starting 
from hydrogen burning up to silicon burning. 
Reactions of $(\alpha,\gamma)$ and $(\alpha, p)(p, \gamma)$ are also 
included. 

We employ the PTD model as the explosion mechanism using the standard
configurations that have been considered in the literature. 
In particular, an initial flame of shape $c3$ is imposed for all the 
simulation models and the propagation of the flame is modeled by the 
standard level-set method (see \citet{Reinecke1999b} for details).
Finally, we also construct the theoretical light curves from our 
simulation data by using the analytical model for SNIa \citep{Arnett1982}, 
which assumes the photon diffusion limit. 
This model takes three input parameters: the ejecta mass $M_{{\rm ej}}$, 
ejecta velocity $v_{{\rm ej}}$ and the nickel mass $M_{{\rm Ni}}$, which 
can be derived from the simulation results.  
We also employ the opacity $\kappa = 0.1$ g$^{-1}$cm$^2$ and the gamma-ray 
deposition function according to \citet{Arnett1982}.

\section{Results}
\label{sec:results}

\subsection{General properties}

In our simulations, the central density of NM is fixed to be 
$\rho_{c({\rm NM)}}=3 \times 10^9\ {\rm g\ cm}^{-3}$ because it is expected 
that the minimum density needed for triggering the thermonuclear 
explosion is about $2 - 5 \times 10^9 {\ \rm g\ cm}^{-3}$
\citep{Iwamoto1999,Woosley1997,LeSaffre2006,Seitenzahl2011}. 
We treat the DM core mass $M_{{\rm DM}}$ as a parameter in the simulations,
and we assume $m_{{\rm DM}} = 1$ GeV unless otherwise noted. 

The properties of four of our typical simulation models are listed in 
Table~\ref{table:TestPTD}, where
$\rho_{c({\rm NM})}$ ($\rho_{c({\rm DM})}$) and 
$M_{{\rm NM}}$ ($M_{{\rm DM}}$) are the central density
and total mass of NM (DM), respectively. $R$ is the initial 
stellar radius. The first model 2D-PTD-3-0-c3 in the table represents a 
standard model without DM. 
The other three models in the table have different DM core masses
ranging from $M_{{\rm DM}}=0.01$ to $0.03 M_\odot$. 
All these configurations are very close to the corresponding 
Chandrasekhar-mass limits for the given $M_{{\rm DM}}$. 
The resulting mass of ${}^{56}$Ni, energy released 
through nuclear reactions $E_{{\rm nuc}}$, and total energy $E_{{\rm tot}}$ are 
also presented in the table.  
For the same central density of NM, the baryonic 
mass of the star decreases as $M_{{\rm DM}}$ increases. However, the (baryonic) 
radius of the star increases with $M_{{\rm DM}}$. 

\begin{table*}
\begin{center}
\caption{Simulation setup for four PTD models: 
central densities of NM $\rho_{ c{\rm (NM)} }$
and DM $\rho_{ c{\rm (DM)} }$ are in units of 
$10^{9}$ g cm$^{-3}$.  
Masses of the baryonic matter $M_{{\rm NM}}$, 
dark matter $M_{{\rm DM}}$, and the final 
nickel-56 mass $M_{{\rm Ni}}$ are in units of solar mass. 
$R$ is the initial stellar radius. 
$E_{\rm nuc}$ and $E_{\rm tot}$ are the energy released by nuclear reactions
and final total energy, respectively, both in units of $10^{50}$ erg. 
We assume $m_{{\rm DM}} = 1$ GeV.}
\begin{tabular}{|c|c|c|c|c|c|c|c|c|}
\hline
Model & $\rho_{c({\rm NM})}$ & $\rho_{c({\rm DM})}$ & 
$M_{{\rm NM}}$ & $M_{{\rm DM}}$ & $R$ (km)& $M_{{\rm Ni}}$ & $E_{{\rm nuc}}$ & $E_{{\rm tot}}$ \\ \hline
2D-PTD-3-0-c3 & 3.0 & 0.0 & 1.377 & 0.00 & $1.92 \times 10^3$ & 0.33 & 7.2 & 2.1 \\
2D-PTD-3-1-c3 & 3.0 & 150 & 1.313 & 0.01 & $2.22 \times 10^3$ & 0.25 & 5.8 & 1.5 \\
2D-PTD-3-2-c3 & 3.0 & 530 & 1.223 & 0.02 & $2.79 \times 10^3$ & 0.14 & 3.9 & 0.46 \\
2D-PTD-3-3-c3 & 3.0 & 1050 & 1.015 & 0.03 & $4.25 \times 10^3$ & $2.9 \times 10^{-2}$ & 1.1 & -0.85 
\label{table:TestPTD} \\ \hline 
\end{tabular}
\end{center}
\end{table*}

\begin{figure}
\centering
\includegraphics*[width=8cm, height=6cm]{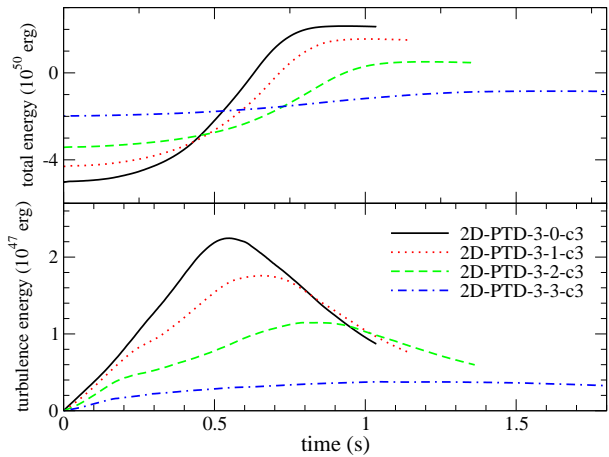}
\caption{Upper panel: total energy against time for the models listed in 
Table~\ref{table:TestPTD}. Lower panel: same as above, but for the total 
turbulence energy.}
\label{fig:model1_2d_PTD_energy}
\end{figure}

In the upper panel of Fig.~\ref{fig:model1_2d_PTD_energy} we plot
the total energy for the models listed in Table~\ref{table:TestPTD}.
Note that for a fixed central NM density, the total NM mass $M_{\rm NM}$ 
decreases as $M_{\rm DM}$ increases. As a result, the initial total energy 
increases with $M_{\rm DM}$ because most of the binding energy is contributed
by NM. 
At early time, models with less DM have faster energy growth than those with 
a more massive DM core. 
The total energy released, by comparing the initial and 
final energies, decreases when $M_{{\rm DM}}$ increases.  
For Model 2D-PTD-3-3-c3, the effects of the admixed DM core are so 
large that the WD remains bound at the end of the simulation due to its 
much lower energy release. 
In the lower panel of Fig.~\ref{fig:model1_2d_PTD_energy}, we plot the total 
turbulence kinetic energy against time for the same models. Similar to the
total energy released, the sub-grid turbulence energy drops when 
$M_{{\rm DM}}$ increases.
The upper panel of Fig.~\ref{fig:model1_2d_PTD_energy} also shows that the 
rate of energy release, reflected by the slopes of the curves, decreases as 
$M_{\rm DM}$ increases, which implies that the time 
needed for a WD to reach the same amount of burnt matter increases. 
We list the mass fractions of major elements at the end of the simulations
in Table \ref{table:Run2D_PTD_results}. In general, the amounts of unburnt 
fuel and IME increase with $M_{\rm DM}$, while those of iron-peaked 
elements 
drop. 
For example, the unburnt ${}^{12}$C are about 34\% and 46\% of the total mass
for models 2D-PTD-3-0-c3 ($M_{\rm DM}=0$) and 2D-PTD-3-3-c3 
($M_{\rm DM}=0.03 M_\odot$), respectively. On the other hand, the mass fraction
of ${}^{56}$Ni decreases significantly from about 24\% to 2.8\% as 
$M_{\rm DM}$ increases from 0 to $0.03 M_\odot$. 
This is related to the different initial density distributions in the 
models. For a larger $M_{{\rm DM}}$, the amount of matter that can reach  
sufficiently high density for complete combustion decreases.


\begin{table*}
\begin{center}
\caption{Mass fractions (normalized by the stellar mass) of all isotopes at 
the end of the simulations for the models listed in 
Table~\ref{table:TestPTD}. }

\begin{tabular}{|c|c|c|c|c|}
\hline
Isotope & 2D-PTD-3-0-c3 & 2D-PTD-3-1-c3 & 
		  2D-PTD-3-2-c3 & 2D-PTD-3-3-c3 \\ \hline
$^{12}$C & 0.30 & 0.32 & 
		   0.36 & 0.43\\
$^{16}$O & 0.33 & 0.35 & 
		   0.39 & 0.46\\
$^{24}$Mg & $1.4 \times 10^{-2}$ & $1.4 \times 10^{-2}$ & 
			$1.4 \times 10^{-2}$ & $1.0 \times 10^{-3}$\\
$^{28}$Si & $5.2 \times 10^{-2}$ & $5.9 \times 10^{-2}$ & 
			$4.9 \times 10^{-3}$ & $3.9 \times 10^{-2}$ \\ 
$^{32}$S & $1.8 \times 10^{-2}$ & $1.9 \times 10^{-2}$ &  
		   $1.6 \times 10^{-2}$ & $1.1 \times 10^{-2}$ \\
$^{36}$Ar & $3.2 \times 10^{-3}$ & $3.1 \times 10^{-3}$ & 
			$2.5 \times 10^{-3}$ & $1.7 \times 10^{-3}$ \\
$^{40}$Ca & $3.1 \times 10^{-3}$ & $2.9 \times 10^{-3}$ & 
			$2.3 \times 10^{-3}$ & $1.3 \times 10^{-3}$ \\
$^{44}$Ti & $6.5 \times 10^{-6}$ & $9.6 \times 10^{-6}$ & 
			$7.3 \times 10^{-6}$ & $2.4 \times 10^{-6}$ \\
$^{48}$Cr & $1.5 \times 10^{-4}$ & $1.4 \times 10^{-4}$ & 
			$8.8 \times 10^{-4}$ & $2.3 \times 10^{-5}$ \\
$^{52}$Fe & $4.8 \times 10^{-3}$ & $3.8 \times 10^{-3}$ & 
			$2.3 \times 10^{-3}$ & $5.1 \times 10^{-4}$ \\
$^{54}$Fe & $5.1 \times 10^{-2}$ & $4.5 \times 10^{-2}$ & 
			$3.3 \times 10^{-2}$ & $1.7 \times 10^{-2}$ \\
$^{56}$Ni & 0.22 & 0.18 & 0.11 & $2.8 \times 10^{-2}$\label{table:Run2D_PTD_results}\\ \hline
\end{tabular}
\end{center}
\end{table*}

\begin{figure}
\centering
\includegraphics[width=8cm, height=6cm]{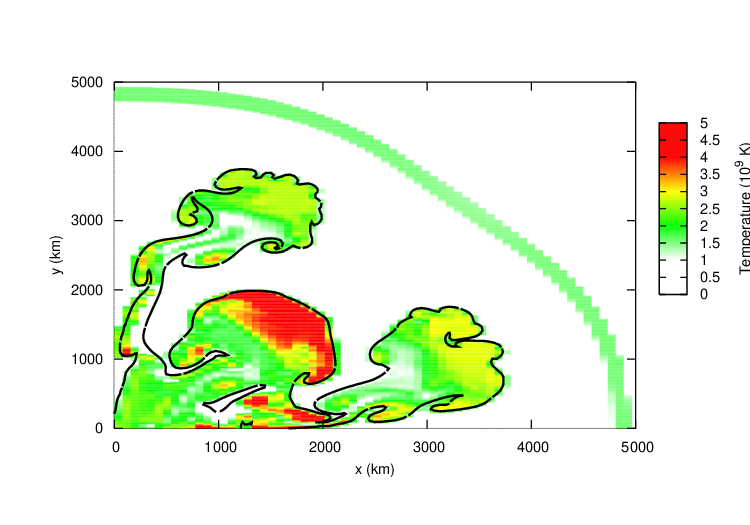}
\caption{The flame surface of Model 2D-PTD-3-0-c3 at $t = 1$ s.}
\label{fig:model1a_2d_PTD_flame8}
\end{figure}

\begin{figure}
\centering
\includegraphics[width=8cm, height=6cm]{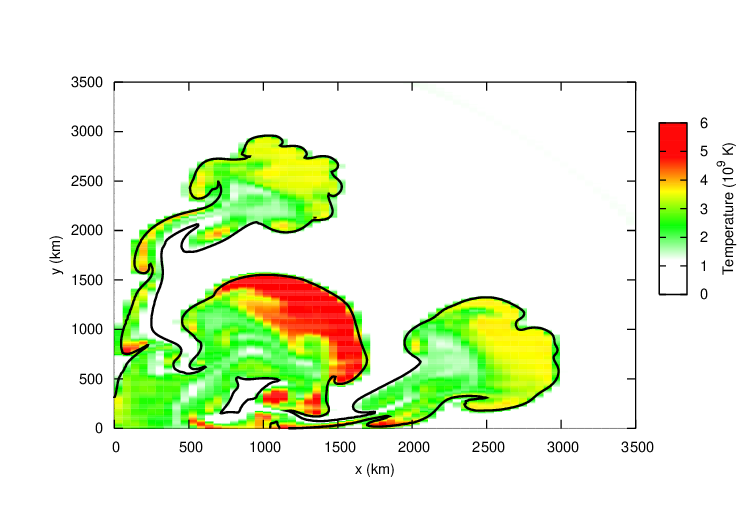}
\caption{Same as Fig. \ref{fig:model1a_2d_PTD_flame8}, but
for Model 2D-PTD-3-1-c3.}
\label{fig:model1b_2d_PTD_flame8}
\end{figure}

\begin{figure}
\centering
\includegraphics[width=8cm, height=6cm]{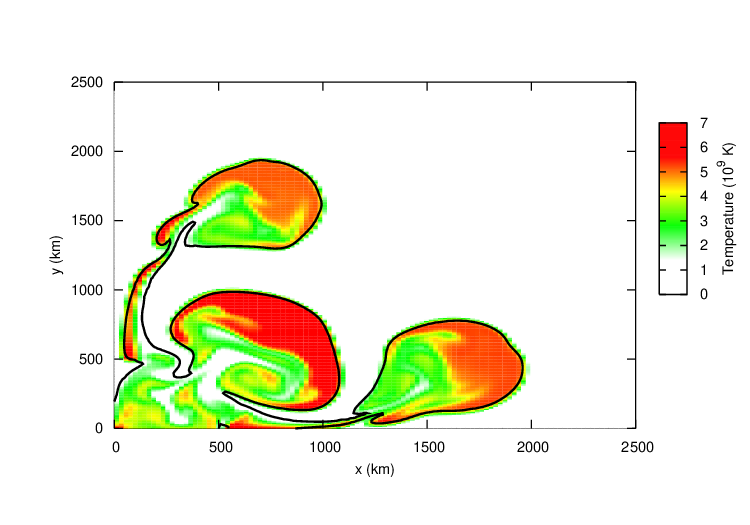}
\caption{Same as Fig. \ref{fig:model1a_2d_PTD_flame8}, but
for Model 2D-PTD-3-2-c3.}
\label{fig:model1c_2d_PTD_flame8}
\end{figure}

\begin{figure}
\centering
\includegraphics[width=8cm, height=6cm]{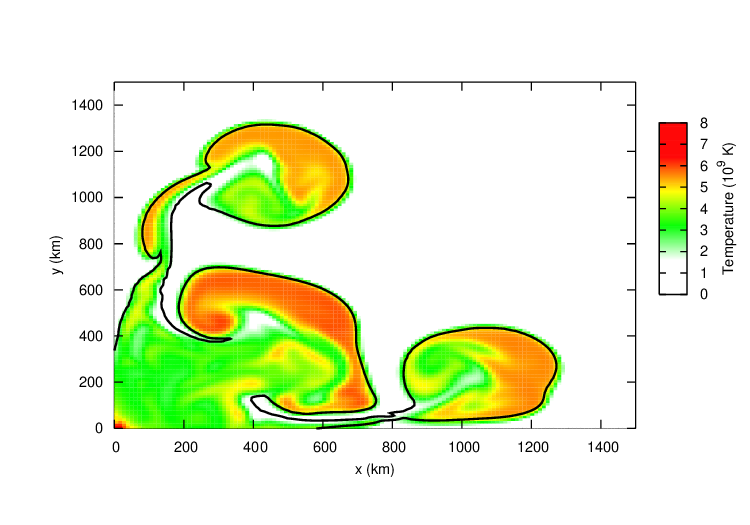}
\caption{Same as Fig. \ref{fig:model1a_2d_PTD_flame8}, but
for Model 2D-PTD-3-3-c3.}
\label{fig:model1d_2d_PTD_flame8}
\end{figure}

Next we consider the effects of DM on the flame surface. We plot in 
Figs.~\ref{fig:model1a_2d_PTD_flame8}-\ref{fig:model1d_2d_PTD_flame8} the 
flame surfaces (represented by the temperature) at $t=1$ s
for the models listed in Table~\ref{table:TestPTD}. 
In Model 2D-PTD-3-0-c3, which is the standard PTD model without DM 
considered in the literature (see for example \citet{Niemeyer1997b}, 
\citet{Reinecke1999b} and \citet{Reinecke2002a}), the flame shows a 
convoluted structure, with clear instabilities of flame-fluid interaction 
including the Rayleigh-Taylor instabilities and Kelvin-Helmholtz 
instabilities. The injection of fuel into the flame can be seen 
as well. 
When $M_{{\rm DM}}$ increases, the injection of fuel can still be found. 
But the Kelvin-Helmholtz instabilities are suppressed. The flame surface also 
becomes smoother and less turbulent.

\subsection{Connection with Sub-luminous Supernovae}


The light curve of a sub-luminous SNIa has a low peak luminosity, suggesting 
that the $^{56}$Ni content is lower than ordinary SNIa. 
The explosion is very weak and only partial ejecta 
are dispelled instead of a disruption of the whole star. 
Notice that, how much and how the mass is ejected in the explosion 
are not clear unless the simulation continues
until the homologous expansion phase is reached, which takes place about ten seconds
after the deflagration/detonation stage has ended 
\citep{Roepke2005a, Roepke2005b}.
However, this involves using either a sufficiently large simulation box which
can accommodate the rapidly expanding ejecta or expanding meshes to 
prevent the ejecta from leaving the box \citep{Roepke2005b}.

Since the amount of ejecta mass is an important parameter in constructing 
the resultant light curves from the simulations, we constrain it 
in the following ways. First the minimum ejecta mass is estimated by counting 
all the fluid elements with positive energy at the end of the simulation. 
Second we assume that the maximum ejecta mass is equal to the total mass of 
the star. 

\begin{figure}
\centering
\includegraphics*[width=8cm, height=6cm]{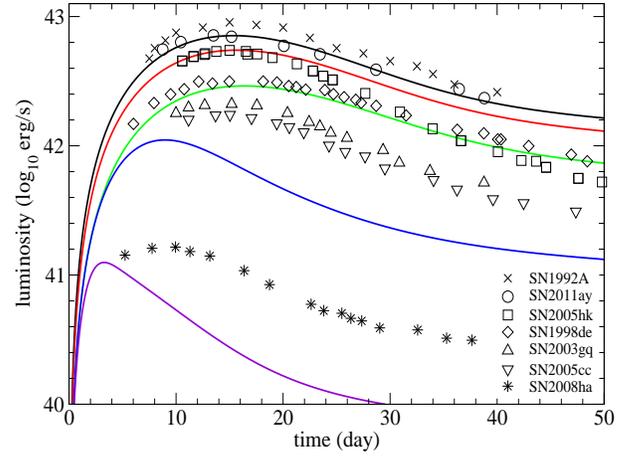}
\caption{From the top solid line to the next-to-bottom solid line:
bolometric light curves of Models 2D-PTD-3-0-c3, 2D-PTD-3-1-c3, 
2D-PTD-3-2-c3 and 2D-PTD-3-3-c3. The lowest light curve corresponds
to an extra model, with similar configurations as the above 
four models but with $M_{{\rm DM}} = 0.032 M_{\odot}$.
Observational data of SN2011ay, SN2005hk, SN1999by, SN2003gq,
SN2005cc and SN2008ha are included.}
\label{fig:model1_2d_PTD_alc_wdata}
\end{figure}

In Fig. \ref{fig:model1_2d_PTD_alc_wdata} we show the bolometric light curves 
for the four simulation models listed in Table~\ref{table:TestPTD} 
(from the top solid line to the next-to-bottom solid line) and also 
one additional model (the bottom line) not listed in the table. 
The extra model has a DM core mass $M_{{\rm DM}} = 0.032 M_{\odot}$. 
For the first three models (2D-PTD-3-0-c3, 2D-PTD-3-1-c3, 2D-PTD-3-2-c3), 
the total final energies are positive and hence we use the maximum ejecta mass
for each model to construct the light curves. On the other hand, we use 
the minimum ejecta mass to construct the light curves for the remaining two 
models because their total final energies are negative. 
Fig.~\ref{fig:model1_2d_PTD_alc_wdata} shows that the peak luminosity 
depends sensitively on $M_{{\rm DM}}$. In particular, it can decrease by almost 
two orders of magnitude as $M_{{\rm DM}}$ increases from 0 to $0.032 M_\odot$. 
In the figure, we also plot the data from the constructed bolometric
light curves for some examples of sub-luminous SNIa for comparison. 
It can be seen that our SNIa simulations with admixed DM give a range 
of peak luminosities that covers the observed sub-luminous SNIa including 
the exceptionally dim SN2008ha.

\begin{figure}
\centering
\includegraphics*[width=8cm, height=6cm]{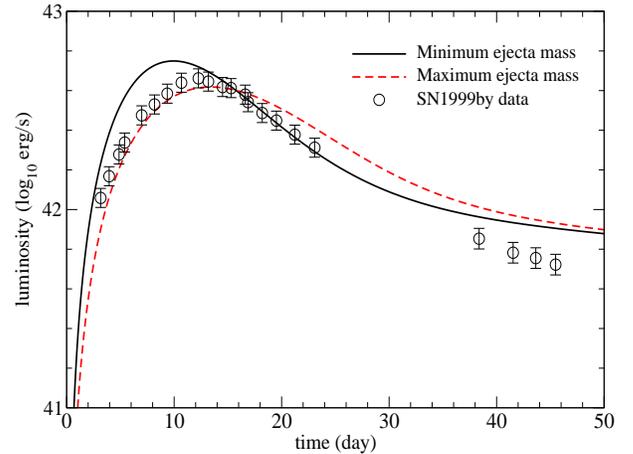}
\caption{Bolometric light curves of a simulation model 
with $M_{{\rm DM}} = 0.02 M_{\odot}$. The solid (dashed) line is 
constructed by assuming that the ejecta mass takes the minimum (maximum) 
value estimated from the simulation as discussed in the text. 
Observational data of SN1999by is also included. The error bars
correspond to the uncertainties in the distance modulus
and the measurements.}
\label{fig:SN1999by}
\end{figure}

\begin{figure}
\centering
\includegraphics*[width=8cm, height=6cm]{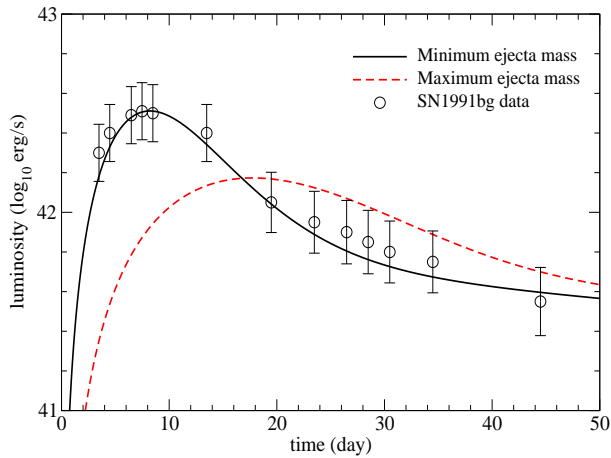}
\caption{Same as Fig. \ref{fig:SN1999by} but for a simulation 
model with $M_{{\rm DM}} = 0.026 M_{\odot}$. Observational data of 
SN1991bg is included for comparison. The error bars correspond to the 
uncertainties in the distance modulus and the measurements.}
\label{fig:SN1991bg}
\end{figure}

Our results suggest that the variations of the observed light curves of 
different sub-luminous SNIa may be due to the fact that the underlying WDs
of the systems contain different amounts of DM. 
For a given observed SNIa light curve, we can use $M_{\rm DM}$ as a 
parameter for performing hydrodynamical simulations to fit the observed data. 
In principle, for a given $M_{\rm DM}$, a unique light curve can be
determined by the resulting velocity profile and total mass of ejecta 
obtained from the simulation. However, as we discussed above, the total mass 
of ejecta cannot be  determined accurately from the simulations due to 
computational limitations. 
We thus calculate two different light curves corresponding to the minimum
and maximum ejecta masses for a chosen $M_{\rm DM}$.

%


In Fig. \ref{fig:SN1999by} we plot the bolometric light curves by using 
the results from a simulation model with $M_{{\rm DM}} = 0.02 M_{\odot}$. 
The solid and dashed lines in the figure are obtained by using the 
minimum and maximum ejecta masses, respectively. This simulation 
model has a positive total final energy. As a result, the minimum and 
maximum ejecta masses estimated from the simulation are comparable, and
hence the two constructed light curves are also quite close to each other. 
We expect that the light curve corresponding to the actual ejecta mass
should lie between the two limits. 
In the figure, the observational data of a sub-luminous supernova SN1999by 
(with error bars) are also presented for comparison. The error bars correspond 
to the uncertainties in the distance modulus and measurements. 
It is seen that the data around the peak luminosity lie very close to the 
region between the two constructed theoretical light curves, which 
represents effectively our uncertainty in the calculation.  
At later time, the observational data decays faster than the theoretical 
light curves. One possible reason may be that our assumption of the 
opacity law and gamma-ray deposition function are no longer valid at later time. 

As a different example, we plot in Fig.~\ref{fig:SN1991bg}
the bolometric light curves of a simulation model with 
$M_{{\rm DM}} = 0.026 M_{\odot}$. Contrary to the simulation model with 
$M_{\rm DM}=0.02 M_{\odot}$ discussed above, this model has a negative total 
final energy and hence the estimated minimum ejecta mass differs from the 
maximum ejecta mass quite significantly. As a result, the two corresponding 
light curves (solid and dashed lines) are not close to each other. 
In the figure, the observational data of another sub-luminous supernova 
SN1991bg is also plotted, which agrees quite well with the 
theoretical light curve constructed with the minimum ejecta mass.

The above examples show that the admixture of dark matter
can produce SNIa light curves with large variations 
in peak luminosities comparable with those of 
sub-luminous SNIa. $M_{{\rm ej}}$ also affects 
the peak luminosity, but its influence is much less pronounced 
than $M_{{\rm DM}}$. On the other hand, $M_{{\rm ej}}$ 
dominates the width of a light curve. This suggests that
the observational data of an SNIa can provide hint on the 
ejecta mass and its admixed DM mass, with
$M_{{\rm DM}}$ determining the peak luminosity
while $M_{{\rm ej}}$ the light curve width.
Given the light curve data of an SNIa, we search for the 
best-fitted pair of $M_{{\rm DM}}$ and $M_{{\rm ej}}$, where 
the values of $M_{{\rm Ni}}$ and $v_{{\rm ej}}$ are
derived from simulations as functions of $M_{{\rm DM}}$. 
Models with the minimum chi-squared values 
are chosen to be the representing models of that SNIa. 
In Table \ref{table:FitResult} we list the models with minimum 
chi-squared values of several well observed sub-luminous SNIa. The 
table lists the agreeing $M_{{\rm ej}}$, $M_{{\rm DM}}$ with 
their implied $M_{{\rm Ni}}$, ejecta velocity $v_{{\rm ej}}$
and the two ejecta mass limits $M_{{\rm ej~(max)}}$ and $M_{{\rm ej~(min)}}$,
which are derived from simulations.
We regard that the DM admixture of this model can be
a possible explanation of an observed 
sub-luminous SNIa if its fitted $M_{{\rm ej}}$
lies within the two limits of $M_{{\rm ej}}$.
All sub-luminous SNIa in the list except 
SN1991bg and SN1993H give an $M_{{\rm ej}}$ 
inside the two limits, showing that these SNIa could possibly
have admixed DM cores in the WD progenitors. 
The SN1993H (SN1991bg) has an ejecta mass just 
above (below) the upper (lower) $M_{{\rm ej}}$ limit,
showing that the observational data is 
declining slower (faster) than what the current 
configuration can provide.

\begin{figure}
\centering
\includegraphics*[width=8cm, height=6cm]{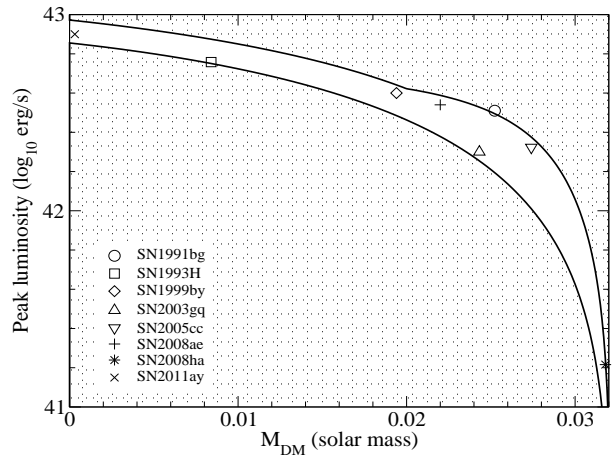}
\caption{The SNIa peak-luminosity against $M_{{\rm DM}}$.
Peak luminosities of some observed SNIa are also plotted. 
The shaded regions are excluded by the mass
bounds of the progenitor mass and minimum ejecta mass
derived from simulations.}
\label{fig:LC_fit}
\end{figure}

In Fig. \ref{fig:LC_fit} we plot the peak-luminosity 
against the fitted $M_{{\rm DM}}$ of the mentioned
SNIa. The peak luminosities of some observed
SNIa are included as data points. The 
shaded regions of the plot are excluded by this
model because these regions correspond to models with 
an ejecta mass out of the bounds. Given an $M_{{\rm DM}}$ the 
range of peak luminosities is very small compared to the observed
range of SNIa peak luminosities. This implies that the admixed
DM mass can be well constrained by the peak luminosity.

\begin{table*}
\begin{center}
\caption{Fitting results of observed 
sub-luminous SNIa. Masses are in units
of solar mass and the ejecta velocity $v_{{\rm ej}}$ 
are in units of $10^9$ cm s$^{-1}$. $M_{{\rm ej~(max)}}$
and $M_{{\rm ej~(min)}}$ are the maximum and 
minimum ejecta masses derived from simulations. 
The last column marks the possibility of using 
admixed DM to explain the observed SNIa.
$m_{{\rm DM}}= 1$ GeV is assumed. }
\begin{tabular}{|c|c|c|c|c|c|c|c|}
\hline
Supernova & $M_{{\rm DM}}$ & $M_{{\rm ej}}$ & $M_{{\rm Ni}}$ & $v_{{\rm ej}}$ & 
$M_{{\rm ej~(max)}}$ & $M_{{\rm ej~(min)}}$ & DM origin \\ \hline
SN1991bg & 0.025 & 0.20 & 0.084 & 0.44 & 1.14 & 0.21 & No \\ 
SN1993H  & 0.008 & 1.37 & 0.216 & 0.74 & 1.35 & 0.69 & No \\
SN1999by & 0.019 & 0.61 & 0.148 & 0.58 & 1.24 & 0.54 & Yes \\
SN2003gq & 0.024 & 0.82 & 0.094 & 4.61 & 1.16 & 0.25 & Yes \\
SN2005cc & 0.027 & 0.33 & 0.058 & 3.70 & 1.01 & 0.14 & Yes \\
SN2008ae & 0.022 & 0.67 & 0.120 & 5.22 & 1.20 & 0.38 & Yes \\
SN2008ha & 0.032 & 0.12 & 0.004 & 2.19 & 1.00 & 0.09 & Yes \\
SN2011ay & 0.000 & 1.35 & 0.320 & 7.63 & 1.38 & 0.73 & Yes\label{table:FitResult} \\ \hline 
\end{tabular}
\end{center}
\end{table*}

In summary, our work shows that the admixture of DM can explain the 
observation data of sub-luminous SNIa. However, as discussed in Sec. 1, 
it should be noted that matching of the bolometric light curves of sub- 
luminous SNIa can also be achieved in other models 
\citep{Pakmor2010, Kromer2013, Kromer2013b, Fink2014, Kromer2015}.

\section{Conclusion}
\label{sec:conclude}

\begin{figure}
\centering
\includegraphics*[width=8cm, height=6cm]{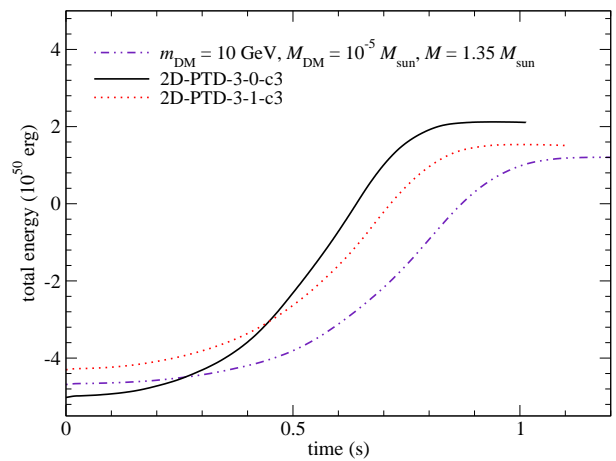}
\caption{Total energy against time for Models 2D-PTD-3-0-c3, 
2D-PTD-3-1-c3 and an extra model similar to 
Model 2D-PTD-3-1-c3 but with $M_{{\rm DM}} = 10^{-5}
M_{{\odot}}$ and $m_{{\rm DM}} = 10$ GeV.}
\label{fig:energy_mDM}
\end{figure}

In this paper, we have performed two-dimensional Newtonian hydrodynamic 
simulations to study the effects of DM on the thermonuclear explosion of 
WDs near the Chandrasekhar mass limit. 
Our initial models are constructed by solving the two-fluid hydrostatic 
equilibrium equations for NM and DM with a fixed central NM density of
$3 \times 10^9 {\ \rm g\ cm}^{-3}$, which is expected to be near the 
minimum density for triggering the explosion. 
The typical models studied by us are solar-mass WDs with small DM cores
($\sim 0.01 M_\odot$) formed by DM with a particle mass of 1 GeV.
As a first step towards understanding the effects of DM on SNIa, we 
assume that the DM core is stationary during the evolution, and we only model the 
dynamics of the NM fluid. This should be a good approximation as the DM core
should be affected mainly by its self-gravity due to its high compactness. 
We employ the PTD model as the explosion mechanism and use the standard 
level-set method to model the flame surface during the dynamical evolution.

We have only considered the PTD model as the explosion mechanism in 
this work. It would be interesting to extend our work by
using other possible explosion mechanisms such as the DDT and GCD models
as discussed in Sec.~\ref{sec:intro}. Finally, we have also assumed that the 
DM core is formed by non-self-annihilating fermionic DM with particle mass 
1 GeV. How would our results be changed if one considers much more massive DM 
particle candidates?  We plot in Fig. \ref{fig:energy_mDM} the total energy
against time for Models 2D-PTD-3-0-c3, 2D-PTD-3-1-c3 and an extra model 
similar to Model 2D-PTD-3-1-c3 but with $M_{{\rm DM}} = 10^{-5}
M_{{\odot}}$ and $m_{{\rm DM}} = 10$ GeV. The initial mass of 
the extra model is 1.35 $M_{\odot}$. Despite the small
$M_{{\rm DM}}$ in the new case, the effects of DM on the 
energy release during explosion is comparable with 
Model 2D-PTD-3-1-c3, which has $M_{{\rm DM}} = 0.01 M_{{\odot}}$.
As found in our previous work \cite{Leung2013}, 
for the same $M_{{\rm DM}}$ or central density, a 
higher $m_{{\rm DM}}$ has a stronger effect on the 
density profile comparing to the case of $m_{{\rm DM}} = 1$ 
GeV. It is because the DM core is more compact and creates
a stronger gravitational attraction field, which 
changes the initial density profile more significantly.  
These changes in the density profiles are also
reflected by the drop of energy release in SNIa explosions.
Our results show that SNIa explosions are sensitive
to both the drop of Chandrasekhar mass, as shown in 
the 1 GeV case, and to the particle mass.
However, modeling more 
massive DM core with $m_{{\rm DM}} = 10$ GeV 
is difficult because the region expected to 
be admixed with DM will be even smaller, implying that
a much higher resolution is needed in order to model
both DM and NM consistently.


Our numerical results show that an increase in $M_{\rm DM}$ or
$m_{{\rm DM}}$ leads to a change in the SNIa explosion
by either decreasing the Chandrasekhar limit for low
$m_{{\rm DM}}$ or altering the density profile for high $m_{{\rm DM}}$. 
First, the explosion becomes weaker and the total energy release is reduced. 
The total turbulence energy, which is an important indicator for the 
PTD model, also decreases. Second, the amounts of unburnt fuel and IME 
increase, while those of iron-peaked elements decrease. 
In particular, the total mass 
of ${}^{56}$Ni depends quite sensitively on $M_{\rm DM}$ and decreases from 
about 0.3 to $0.03 M_\odot$ as $M_{\rm DM}$ increases from 0.01 to 
$0.03 M_\odot$ for the $m_{{\rm DM}} = 1$ GeV case. Finally, the Kelvin-Helmholtz 
instabilities are suppressed and the flame 
surface also becomes smoother and less turbulent as $M_{\rm DM}$ increases. 
We have also constructed the bolometric light curves from our simulations
and compared them with the observational data of sub-luminous 
SNIa. Our results shows that varying the DM core mass from about 0.01 to 
$0.03 M_\odot$ yields a range of peak luminosities that covers the 
observational data very well. The variations of the observed 
light curves of different sub-luminous SNIa may be due to the fact that the 
precursor WDs contain different amounts of DM.

\section{Acknowledgment}
\label{sec:ack}

We thank F. X. Timmes 
for his open-source subroutines for the Helmholtz
equation of state, nuclear reaction network and
the neutrino emission luminosity. 
This work is partially supported by a grant from the
Research Grant Council of the Hong Kong Special
Administrative Region, China (Project No. 400910)
and a CUHK Direct Grant 4053069. SCL is supported 
by the Research Grant Council of the 
Hong Kong Government through the Hong Kong
PhD Fellowship Scheme. 

%
%

\bibliographystyle{apj}
\pagestyle{plain}
\bibliography{biblio}

\section{Appendix: Effects of DM Admixture on SNIa progenitors}

\begin{table*}
\begin{center}
\caption{Stellar properties at the end of 
simulations and main-sequence lifetime 
for models with $M_{{\rm NM}} = 4$
or $7 M_{\odot}$. $M_{{\rm He}}$ and $M_{{\rm CO}}$
are the masses of ${^4}$He, $^{12}$C
and $^{16}$O in units of solar mass. $X_C$
and $X_O$ are the mass fractions of 
$^{12}$C and $^{16}$O of the innermost 
mass shell. $T_{{\rm H~start}}$ ($T_{{\rm H~end}}$) and
$T_{{\rm He~start}}$ ($T_{{\rm He-end}}$) are the 
beginning (ending) time for hydrogen
amd helium burning in units of years.}
\begin{tabular}{|c|c|c|c|c|c|c|c|c|c|}
\hline
$M_{{\rm NM}}$ & $M_{{\rm DM}}$ & $M_{{\rm He}}$ & $M_{{\rm CO}}$ & 
$X_C$ & $X_O$ & $T_{{\rm H~start}}$ & $T_{{\rm H~end}}$ & $T_{{\rm He~start}}$ & $T_{{\rm He~end}}$ \\ \hline
4 & 0 & 0.24 & 0.18 & 0.40 & 0.58 & $1.12 \times 10^6$ & $1.49 \times 10^8$ & $1.55 \times 10^8$ & $1.89 \times 10^8$ \\
4 & 0.01 & 0.24 & 0.19 & 0.47 & 0.50 & $1.08 \times 10^6$ & $1.46 \times 10^8$ & $1.50 \times 10^8$ & $1.82 \times 10^8$ \\
4 & 0.02 & 0.18 & 0.37 & 0.56 & 0.42 & $1.05 \times 10^6$ & $1.44 \times 10^8$ & $1.47 \times 10^8$ & $1.82 \times 10^8$ \\
4 & 0.03 & N/A & N/A & N/A & N/A & $1.01 \times 10^6$ & $1.44 \times 10^8$ & $1.47 \times 10^8$ & N/A \\
7 & 0 & 0.32 & 0.13 & 0.44 & 0.54 & $2.30 \times 10^5$ & $3.98 \times 10^7$ & $4.05 \times 10^7$ & $4.70 \times 10^7$ \\
7 & 0.01 & 0.31 & 0.13 & 0.44 & 0.54 & $2.30 \times 10^5$ & $3.98 \times 10^7$ & $4.05 \times 10^7$ & $4.70 \times 10^7$ \\
7 & 0.02 & 0.31 & 0.13 & 0.47 & 0.50 & $2.20 \times 10^5$ & $3.92 \times 10^7$ & $3.96 \times 10^7$ & $4.57 \times 10^7$ \\
7 & 0.03 & N/A & N/A & N/A & N/A & $2.19 \times 10^5$ & $3.94 \times 10^7$ & N/A & N/A 
\label{table:MainSeq} \\ \hline 
\end{tabular}
\end{center}
\end{table*}

In this article we have studied how the gravity
of DM affects the explosion energetics of SNIa. 
We have shown that 
with an admixed DM core with a total mass in the order
of $\sim 10^{-2} M_{\odot}$, the $^{56}$Ni production
can be significantly suppressed and the corresponding
light curves are comparable with those of 
sub-luminous SNIa. However, it remains unclear 
whether such a DM admixture can leave observable 
consequences already during the main-sequence 
phase, which is well constrained by observational 
data. Therefore, it is necessary to check if stars 
with DM admixture have unusual evolution paths 
that are inconsistent with observational data, and 
if the chemical compositions of the resultant 
white dwarfs are different from those of conventional cases. 

A star acquires DM mainly by accretion through DM-NM 
scattering or by its inherent admixture that exists
already during its formation stage, where DM acts as
a stellar seed. However, following \citep{Kouvaris2008}
to estimate the DM accretion rate, using conventional DM 
parameters, the typical DM accretion rate is insignificant 
compared with the original mass of the star, even when 
we consider a duration of cosmological timescale. Therefore, it 
is unlikely that a star can acquire DM with a mass
comparable with the host simply by accretion. We 
thus focus on DM which acts as a stellar seed. 
In that case, the gravity of DM is important 
even in the protostellar phase.

We performed main-sequence star simulations by
using an open-source stellar evolution code MESA
(Modules for Experiments in Stellar Astrophysics) 
\citep{Paxton2011, Paxton2013, Paxton2015}, which 
can follow the evolution of a star from the 
protostellar phase up to the white dwarf stage. 
We used the MESA code version 3372, which solves the 
fully coupled one-dimensional structure 
and composition equations simultaneously,
using the Helmholtz EOS to describe
the thermodynamics properties of NM. 
The DM is assumed to be in hydrostatic equilibrium, 
and to a good approximation, the DM profile remains 
static during the simulation. We observed that
due to the compactness of the DM core, in most 
of the stellar lifetime, the DM core has a size 
smaller than the outer radius of the innermost
fluid elements. Effectively, we modified the 
hydrostatic equation in the MESA code
by including the DM core which behaves 
like a point-mass as
\begin{equation}
\frac{dP_{{\rm NM}}}{dm} = - \frac{G (m_{{\rm NM}}(r) + M_{{\rm DM}})}{4 \pi r^4}.
\end{equation}
All notations have the same meaning as those in the main text.
Due to the singular behavior of the DM potential
near the core, there are numerical difficulties that the 
results are resolution dependent. Also, the typical time-step 
becomes prohibitively small, due to the large potential gradient,
which leads to a large density gradient and hence a large
chemical composition gradient near the core. Also, the 
$1/r$ potential from the DM leads to 
divergence in constructing the initial model. 
To ameliorate these problems, we smear out the 
effect of DM by increasing the innermost
fluid elements from $10^{-8}$ to $10^{-4} M_{\odot}$.
This allows us to capture the effects of DM's gravity within
a reasonable simulation time.

\begin{figure}
\centering
\includegraphics*[width=8cm, height=6cm]{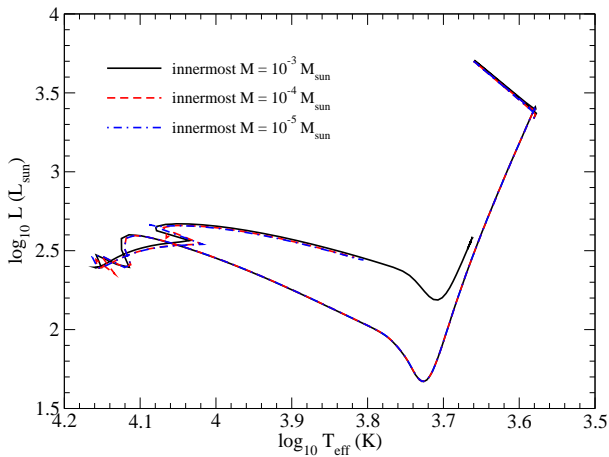}
\caption{The H-R diagrams of stars with 4 $M_{\odot}$ NM and
0.03 $M_{\odot}$ DM but with different innermost mass
shell. The simulation is done until the time-step becomes
smaller than $10^1$ years.}
\label{fig:NM4_dmmin_HR}
\end{figure}

We use the star evolution model 1M\_pre\_ms\_wd
in the {\it test suite} package to follow the 
stellar evolution from the protostellar phase. 
We considered star models with a mass from 4 to 7 
$M_{\odot}$, which are believed to be the 
progenitors of carbon-oxygen white dwarfs. 

\begin{figure}
\centering
\includegraphics*[width=8cm, height=6cm]{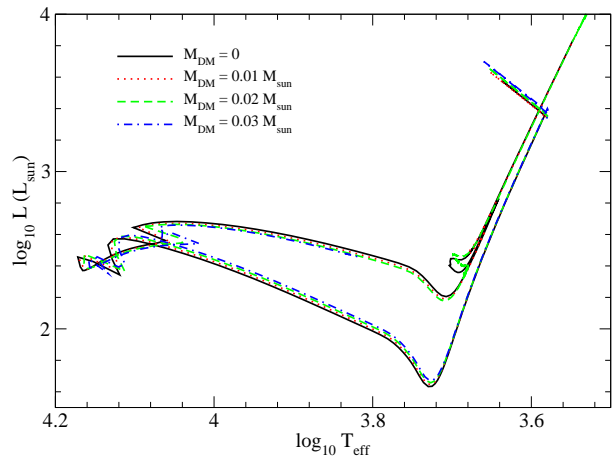}
\caption{The H-R diagrams of stars with 4 $M_{\odot}$ NM. 
$M_{{\rm DM}}$ ranges from 0 to 0.03 $M_{\odot}$.
All simulations have the innermost mass shell
fixed at $10^{-4} M_{\odot}$, and 
they are stopped when the time-step becomes
smaller than $10^1$ years.}
\label{fig:NM4_MDM_HR}
\end{figure}

\begin{figure}
\centering
\includegraphics*[width=8cm, height=6cm]{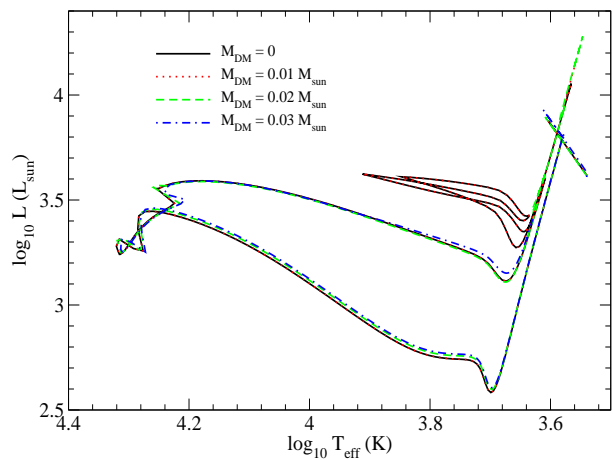}
\caption{Same as Fig. \ref{fig:NM4_MDM_HR}, but 
for $M_{{\rm NM}} = 7 M_{\odot}$.}
\label{fig:NM7_MDM_HR}
\end{figure}

To show that the rise of innermost fluid element mass
does not introduce spurious results, we plot in 
Fig. \ref{fig:NM4_dmmin_HR} the HR diagram of a
star with 4 $M_{\odot}$ NM and 
0.03 $M_{\odot}$ DM, but with different 
innermost fluid element masses. We do not follow the whole
evolution till the formation of the white dwarf because
the timestep has already become prohibitively small
when it enters the helium burning phase. We stopped
the simulation when the average time step drops below
$10^1$ years. In the mass range considered, the qualitative behavior
of the stellar evolution remains unchanged. This shows 
that in this resolution the basic properties of the 
main-sequence phase are captured. The model with 
a higher innermost mass shell can run longer due to 
the stronger smearing of the DM point-mass gravity.

We plot in Figs. \ref{fig:NM4_MDM_HR} and \ref{fig:NM7_MDM_HR}
the HR diagrams for star models with a normal matter mass
of 4 and 7 solar masses, but for different $M_{{\rm DM}}$.
In both figures, the evolution paths of the hydrogen 
burning phase and the helium burning phase are insensitive
to $M_{{\rm DM}}$. We terminated the simulations for 
$M_{{\rm DM}} = 0.03 M$ before the exhaustion of core helium
because of the small time-steps. One qualitative difference
that can be observed is the disappearance of the horizontal branch 
during helium burning for the model with $M_{{\rm NM}} = 7 M_{\odot}$
and $M_{{\rm DM}} = 0.02 M_{\odot}$.

In Table \ref{table:MainSeq} we tabulate the stellar
properties extracted from profiles at the end of simulations,
and also the age of the star when hydrogen or helium 
burning starts or ends. No results are listed for models
with $M_{{\rm DM}} = 0.03 M_{\odot}$ because the 
simulations are terminated before the helium burning 
phase commences. For models with 
$M_{{\rm NM}} = 4 M_{\odot}$, when $M_{{\rm DM}}$ increases, 
the helium mass decreases while the carbon-oxygen
mass increases. Also, hydrogen burning begins
and ends sooner, with the whole
hydrogen burning lifetime shortened. 
Similar features are observed for the
helium burning. The $^{12}$C mass fraction 
increases while that of $^{16}$O decreases. 
The effects of DM in models with 
$M_{{\rm NM}} = 7 M_{\odot}$ become smaller
so that there is almost no change when 
$M_{{\rm DM}} = 0.01 M_{\odot}$. But as
$M_{{\rm DM}}$ further increases, similar effects
can still be observed, including a lower
helium mass, earlier hydrogen and helium burning
and a shorter main-sequence lifetime. Also, an
increase (a decrease) in $^{12}$C ($^{16}$O)
mass fraction is observed.

From the above comparison, we have shown that 
in the mass range of $M_{{\rm DM}}$ considered in the 
main text, the DM core which is assumed to exist 
as early as the star forms, does not alter 
the stellar evolution significantly. Specifically,
all models predict a path during the main-sequence
phase in the HR diagram comparable with the cases 
without DM. Moreover, the final chemical composition of the 
carbon-oxygen white dwarf does not deviate significantly
from what we have assumed, a 50 \% carbon and 50 \% 
oxygen by mass.

\end{document}